# Design of a Resistive Plate Chamber using additive manufacturing

L. Benussi[a], S. Bianco[a], R. Campagnola[a,e], M. Caponero[a,b], S. Colafranceschi[c], H. Gebremedhin[c], J. Hess[c], J. Horsley[c], M. Landis[c], S. Meola[a], D. Nester[c], L. Passamonti[a], L. Peachey-Stoner[c], R. Peachey-Stoner[c], D. Piccolo[a], D. Pierluigi[a], A. Russo[a], G. Saviano[a,d], L. Stutzman[c], R. Tezazu[c]

[a]*Istituto Nazionale di Fisica Nucleare Laboratori Nazionali di Frascati Frascati Italy. ROR: 049jf1a25*
[b]*ENEA Frascati Italy. ROR:01026pq66*
[c]*Eastern Mennonite University Harrisonburg VA 22802 USA ROR:059xmmg10*
[d]*Sapienza Universita' di Roma Rome Italy. ROR:02be6w209*
[e]*Corresponding author and presenter*

**Abstract**

Driven by the recent improvement in additive manufacturing technologies, we designed a detector that can be fully printed with a standard and commercial 3D printer. The main goals of this research concern the marginal design and construction costs, the reproducibility/modularity of the products, and the reduced assembly time. During the first phase of this research, after determining the most suitable material, we produced 10 examples of detector.

## 1. Introduction

Additive Manufacturing (AM) could represent a significant improvement in the building phase of detectors suitable for several applications and future experiments. We propose a novel approach in the construction and characterization of novel detectors using AM. The 3D printing technique, although it may be crucial for High Energy Physics (HEP), is still not fully adopted in detector construction, and not much literature concerning totally-AM-produced detectors is available. The AM technology can decrease construction cost, can be useful in controlling time line and design optimization. We used the AM technology to build small samples of Resistive Plate Chamber (RPC) detectors. RPC's [1] are widely used in HEP as gas detectors for muon particles.

## 2. Detector

To build the detector prototypes for RPC's with the AM novel technique, a custom made filament was introduced [2]. Most of the RPC resistive plates are nowadays built with Bakelite [3], that presents some issues: high construction cost for a single detector, hygroscopy, limited chemical stability [4]. To replace Bakelite, graphene-doped PLA (polylactid acid) was chosen. PLA is a common biodegradable plastic economically produced from renewable resources, one of the most popular materials used in desktop 3D printing. This plastics is commonly used even in desktop printers as the extrusion temperature is relatively low at 200°C, it is environmental-friendly, non toxic and with adequate mechanical properties to be used as the active part of a detector. PLA itself is an insulator and it will not work as an electrode. To prepare the filament and control its resistivity, the PLA was doped with graphene. A wide range of electrical resistivity can be obtained, by varying the doping concentration. Graphene revealed to be an effective option to modify the PLA electrical properties. Figure 1 shows the concentration of graphene as a function of surface resistivity.

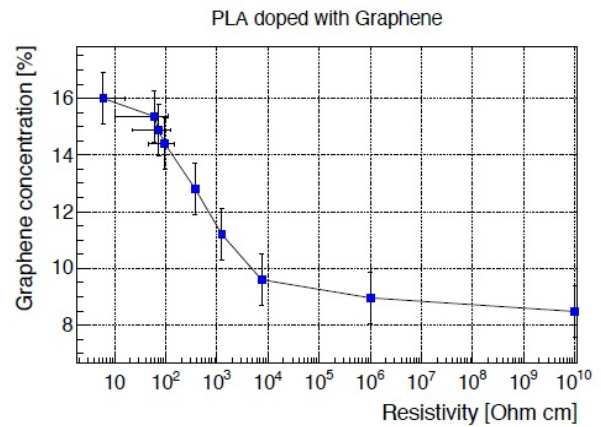

Figure 1: Graphene concentration versus resistivity

The mixed graphene-PLA pellet is then extruded to generate a plastics filament to be used in the 3D printer. The printing phase was performed on glass to ensure a mirror-like surface of the electrode. Figure 2 shows a typical 3D printed graphene-PLA electrode.

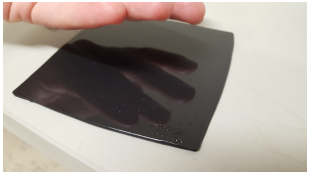

Figure 2: Surface of the 3D printed electrode

Two resistive electrode and one insulating pure PLA frame were printed and glued to form the detector stack and ensure gas tightness. Figure 3 shows a cross-sectional schematic drawing of the detector assembly.

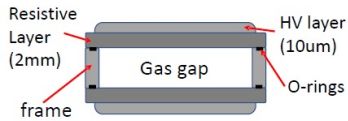

Figure 3: Detector schematic drawing

Finally, a 500 $\Omega/\square$ HV distribution layer, made of graphite powder, is glued on both electrodes to apply HV. Figure 4 shows the first assembled detector, including the gas mixture inlet and outlet.

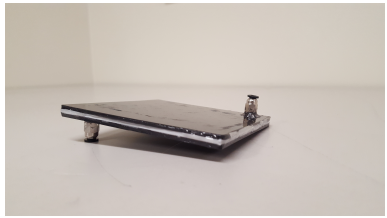

Figure 4: First assembled detector prototype

## 3. Results

Ten detectors were printed and assembled to assess consistency and reproducibility of this technique. A resistivity of order $5 \cdot 10^{10}$ $\Omega$/cm was reached, same order of Bakelite, but new prototypes are easier to build presenting a high reproducibility, no water absorption and are more stable over time. Preliminary test campaign about radiation hardness was conducted; results show the prototype behaviour is consistent with PLA-radiation interaction [5]. Studies about compatibility with environment-friendly gas are ongoing, in particular about compatibility with fluorine-based gas [6] . For the data taking campaign, an Ar/$CO_2$ 70:30 mixture was used.

Moreover, interactions with minimum ionizing particle (MIP) show signals that are order 100 mV, at 4.5 kV with Ar/$CO_2$ 70:30. A data-taking campaign to characterize the detector is being carried out to study signal properties using a dedicated ADC/TDC and a data acquisition system. The ten printed detectors are going to be arranged in a cosmic telescope for general purpose studies.

## 4. Conclusions

Additive Manufacturing allows complete control of the costs, timeline and design optimization. Moreover, prototyping cost and the need for an industrial partner can be reduced by using this novel approach .

Traditional limitations of detector shape is bypassed by the unconstrained 3D printer, which allows curves surfaces or customized form factors. Although commercial printers are still lacking smooth surface printouts, this is a process that can be performed after the actual construction phase. The existing limitation is the actual printable size, the adopted desktop printer was limited to 50x50 $cm^2$ while industrial printer can exceed 100x100 $cm^2$.

Featuring a simple method to control the resistivity, simply varying the amount of graphene, allows to tune the detector and signal properties to match the kind of experiment and front-end electronics. While the entire project design, mechanical stability and construction stay the same, the electrical properties can be tuned during to match any requirement.

RPC prototyping and test phases will continue with cosmic ray and/or radiation sources. Upon success, several detector concepts with Additive Manufacturing will be studied, such as Thick Gem (THGEM) [7] and Microdot detectors [8] .

## 5. Acknowledgement

This work is supported by the Jeffress Trust Awards Program in Interdisciplinary Research.